\documentclass[twocolumn]{bmcart}

\usepackage[utf8]{inputenc} 
\usepackage{hyperref}       
\usepackage{microtype}      

\usepackage{graphicx}

\usepackage{makecell}

\usepackage{booktabs}

\usepackage[T1]{fontenc}

\usepackage{xcolor}
\usepackage[normalem]{ulem}

\usepackage{colortbl}

\startlocaldefs
\newcommand{\old}[1]{}
\newcommand{\oldd}[1]{}

\endlocaldefs

\begin{document}

\begin{frontmatter}

\begin{fmbox}
\dochead{Software}


\title{DiviK: Divisive intelligent K-Means for hands-free unsupervised clustering in big biological data}


\author[
   addressref={polsl,ng},             
   corref={polsl},                    
   email={Grzegorz.Mrukwa@polsl.pl}   
]{\inits{GM}\fnm{Grzegorz} \snm{Mrukwa}}
\author[
   addressref={polsl},
   email={Joanna.Polanska@polsl.pl}
]{\inits{JP}\fnm{Joanna} \snm{Polanska}}


\address[id=polsl]{
  \orgname{Silesian University of Technology, Department of Data Science and Engineering}, 
  \street{Akademicka 16},                     %
  \postcode{44-100}                           
  \city{Gliwice},                             
  \cny{Poland}                                
}
\address[id=ng]{%
  \orgname{Netguru},
  \street{Ma\l{}e Garbary 9},
  \postcode{61-756}
  \city{Poznań},
  \cny{Poland}
}


\begin{artnotes}
\end{artnotes}



\begin{abstractbox}

\begin{abstract} 
\parttitle{Background}
Investigating molecular heterogeneity provides insights into tumour origin and metabolomics. The increasing amount of data gathered makes manual analyses infeasible - therefore, automated unsupervised learning approaches are utilised for discovering tissue heterogeneity. However, automated analyses require much experience setting the algorithms' hyperparameters and expert knowledge about the analysed biological processes. Moreover, feature engineering is often needed to obtain valuable results because of the numerous mass channels measured.

\parttitle{Methods}
We propose DiviK: a scalable stepwise algorithm with local data-driven feature space adaptation for segmenting high-dimensional datasets. The algorithm is compared to the optional solutions (regular k-means, spatial and spectral approaches) combined with different feature engineering techniques (None, PCA, EXIMS, UMAP, Neural Ions). Three quality indices: Dice Index, Rand Index and EXIMS score, focusing on the overall composition of the clustering, coverage of the tumour region and spatial cluster consistency, are used to assess the quality of unsupervised analyses. Algorithms were validated on two datasets acquired by Mass Spectrometry Imaging – 2D human cancer tissue samples and 3D mouse kidney images.

\parttitle{Results}
DiviK algorithm performed the best among the four clustering algorithms compared (overall quality score 1.24, 0.58 and 162 for $d(0,0,0)$, $d(1,1,1)$ and the sum of ranks, respectively), with spectral clustering being mostly second. Feature engineering techniques impact the overall clustering results less than the algorithms themselves (partial $\eta^2$ effect size: 0.141 versus 0.345, Kendall's concordance index: 0.424 versus 0.138 for $d(0,0,0)$). 

\parttitle{Conclusions}
DiviK could be the default choice in the initial exploration of Mass Spectrometry Imaging data. Thanks to its unique, GMM-based local optimisation of the feature space and deglomerative schema, DiviK results do not strongly depend on the feature engineering technique applied and can reveal the hidden structure in a tissue sample. Additionally, DiviK shows high scalability, and it can process at once the big omics data with more than 1.5 mln instances and a few thousand features. Finally, due to its simplicity, DiviK is easily generalisable to an even more flexible framework. Therefore, it is helpful for other -omics data (as single cell spatial transcriptomic) or tabular data in general (including medical images after appropriate embedding). A generic implementation is freely available under Apache 2.0 license at \url{https://github.com/gmrukwa/divik}.

\end{abstract}


\begin{keyword}
\kwd{machine learning}
\kwd{unsupervised clustering}
\kwd{feature engineering}
\kwd{local feature space adaptation}
\kwd{high-dimensional data analysis}
\kwd{omics}
\kwd{mass spectrometry imaging}
\kwd{tumour heterogeneity}
\end{keyword}


\end{abstractbox}
\end{fmbox}

\end{frontmatter}



\section*{Background}

Mass Spectrometry Imaging (MSI) is widely used for discovering molecular profiles, as it provides unparalleled insight into the metabolomics of tissue samples \cite{aichler2015maldi,miura2010ultrahighly,hattori2010paradoxical}. Applied to a tumour specimen, MSI allows investigating heterogeneities that could indicate functional differences across tissue regions, as well as varying cancer subtypes that require dedicated treatment \cite{djidja2010novel,morita2010imaging,groseclose2008high,quaas2013maldi,steurer2013maldi,pietrowska2017molecular,martinez2017cancer}.

Technically, MSI is an excellent example of big biological data, with the following characteristics:

\begin{itemize}
  \item \emph{Volume}: the spatial resolution of $5-100 \mu m$ leads to 10,000 - 4,000,000 spectra potentially acquired from a $1cm^2$ tissue sample with Time-of-Flight (ToF) spectrometers and becomes an even more severe problem for 3D MSI data \cite{vos2020experimental};
  \item \emph{Velocity}: the last three years have brought more than 4,500 MSI datasets uploaded to the METASPACE database \cite{palmer2017fdr,metaspace2020summary};
  \item \emph{Variety}: a dataset may consist of more than 200,000 mass channels (features) per single spectrum (observation), representing proteins, peptides, and/or metabolites;
  \item \emph{Veracity}: the acquisition methods and various biological phenomena introduce heavy duplication of information, thus capturing the most relevant nuances becomes complicated with dominating high-level patterns amplified \cite{polanski2015signal}. On the other hand, feature importance may be diverse across separate regions of interest.
\end{itemize}

Efficient analyses of thousands of MSI spectra usually require careful feature space adaptation, irrespective of the extensive preprocessing pipeline \cite{jones2011multiple,thomas2016dimensionality,veselkov2014chemo,verbeeck2020unsupervised}. In a fully unsupervised setup, there are the following groups of methods: filtering, linear (e.g. Principal Components Analysis), and non-linear (e.g. Universal Manifold Approximation and Projection) \cite{postma2009dimensionality}. Filtering removes features according to some threshold (often constant). Linear methods produce a new set of features, a linear combination of the input features. Non-linear methods aim to find a low-dimensional representation that best approximates distances between pairs of points. All these methods are applied once globally to the entire dataset.

Many clustering methods are known, and one can divide them into a few significant groups based on their approach: centroid-based (e.g. k-Means), connectivity-based (e.g. hierarchical clustering), distribution-based (e.g. Gaussian Mixture Modeling), and density-based (e.g. graph-cuts clustering). In addition, sometimes, a method is proposed that exercises more than a single property of the data (e.g. both connectivity and centroids) and such a method is called hybrid clustering.

The clustering quality can be captured via metrics like Dunn's index \cite{dunn1974well} or Adjusted Rand Index \cite{lawrence1985comparing}. These were introduced for low-dimensional data clustering, but studies \cite{lipor2020clustering,rodriguez2019clustering} compare their usefulness for high-dimensional data and subspace clustering. Hundreds of features that clustering algorithms process are already seen as a high-dimensionality problem, but it is still at least an order of magnitude less severe than in MSI data analyses. Therefore domain-related work must be investigated \cite{verbeeck2020unsupervised}.

The earliest solutions for Mass Spectrometry Imaging take advantage of Principal Components Analysis (PCA), and agglomerative clustering \cite{deininger2008maldi}. PCA transforms the data, and components explaining 70\% of the variance are selected. The authors use the Euclidean metric to capture spectra similarity and Ward linkage to provide the most meaningful results. The approach is semi-supervised, as it requires the manual setting of the number of clusters based on histological examination.

Another method features high-dimensional data clustering (HDDC) \cite{bouveyron2007hddc} -- a hybrid approach based on the Gaussian Mixture Model (GMM). The fitting process yields locally relevant features, so the domain is independently adapted for each estimated component. The observations' similarity is considered Euclidean metric in some subspaces of the original feature space. For MSI data, a combination of HDDC with edge-preserving denoising of $m/z$-images is applied \cite{alexandrov2010spatial}. Oppositely to smoothing the clustering results, the authors propose an approach oriented on $m/z$ images, so the denoising provides better quality data for the clustering method. The idea of denoising is extended \cite{alexandrov2011efficient}. The Fast Map algorithm adapts the domain, and K-Means clustering follows, which is more straightforward than HDDC. The denoising stage contains the main advancements, which we will omit here. The overlap of clusters and actual structures is the primary quality metric. However, both HDDC and K-Means approaches \cite{alexandrov2010spatial,alexandrov2011efficient} also compare the visual perception of cluster consistency.

Region consistency is the main idea behind EXIMS \cite{wijetunge2015exims} -- a modern PCA-oriented feature engineering approach. First, histogram equalisation enhances the contrast of a molecular image. Then, the algorithm proposes a scoring method to assess whether structures exist in the enhanced molecular image. The random distribution of counts over the image leads to a low score value. Next, typical structure types like regions, curves, gradients and islets are recognised with a grey level co-occurrence matrix, resulting in a higher score value. However, the definition of the score yields unbounded values. Thus there is no clear threshold that can help discern informative peaks. Finally, the informative peaks are PCA-transformed and clustered with the fuzzy C-Means algorithm.

K-Means and spectral clustering are further investigated in a two-step scenario to speed up computations \cite{dexter2017two}. In the first step of clustering, both methods consider several random subsets of input data independently, and then cluster representatives are grouped to form the final clusters. Algorithms use cosine distance to measure spectrum similarity. Authors manually perform hyperparameter selection (i.e. selecting the number of eigenvectors of the connectivity graph) to provide the best results (visually). Both methods isolate 7 clusters per prior settings.

Feature engineering techniques also address the increased dataset volume. For example, Hierarchical Stochastic Neighbor Embedding (HSNE) has been claimed to be superior to classical non-linear techniques \cite{abdelmoula2018interactive}. Reduced feature space is constructed hierarchically, with the approach \emph{Overview-First, Details-on-Demand}. Firstly, characteristic points of the dataset (landmarks) are embedded to provide an overview. Then, the algorithm creates a new embedding for local neighbourhoods. Finally, each point is assigned the likelihood of being represented by a specific landmark. The authors discuss using maps of the likelihood for molecular segmentation given a landmark. However, they propose clusters of landmarks manually.

Another scalable feature engineering example is the Universal Manifold Approximation and Projection (UMAP) algorithm. It is more scalable than t-SNE and preserves more of the global structure \cite{mcinnes2018umap}. Moreover, no restriction is made on the embedding dimensions. Thus, it may be used as a more general feature engineering method in MSI than the visualizing-only capabilities of t-SNE \cite{smets2019evaluation}.

The recent years introduced a new category of unsupervised segmentation approaches based on deep learning. They form groups that correspond to dual interpretation of MSI data:

\begin{itemize}
  \item \emph{Spectrum oriented}: neural network creates a low-dimensional embedding of spectra \cite{inglese2017deep,gardner2019visualizing,abdelmoula2021peak} that is further analysed with classical clustering methods or visualised.
  \item \emph{Ion-image oriented}: molecular images are analysed to find similarities in molecule distribution and enhance biologically relevant information. The neural approach extends the idea from EXIMS \cite{wijetunge2015exims} and classical community detection algorithms \cite{wullems2019detection}. It mimics a human recognising detailed molecular patterns and clusters the ion images based on spatial expressions \cite{zhang2021spatially}. This way, new molecular features are obtained, simultaneously representing whole groups of ion images. These \emph{neural ion images} are subject to a classical unsupervised analysis pipeline, including pixel-wise approaches.
\end{itemize}

The unsupervised analyses of MSI data via clustering outlined above follow a single schema: \emph{some} domain adaptation is applied globally, and a basic clustering algorithm is applied. This approach has several drawbacks.

\begin{itemize}
  \item PCA applied to the whole dataset can capture the most variance on a high level, but at the same time, it discards all nuances. These nuances might be crucial for hierarchical analyses. Other domain adaptation methods may partially resolve the issue, but the disproportion between low- and high-level detail is significant with thousands of dimensions.
  \item Non-hierarchical approaches (e.g. one-step, two-step) cannot provide insights into the internal structures unless the molecular differences dominate the whole dataset.
  \item Only hierarchical approaches describe the relation of subgroups as a clustering tree that captures context like \emph{parent}, \emph{child}, or \emph{sibling}. This context provides additional insight into molecular diversity. For example, the parent-child relation of clusters could describe different functional subregions of the tumour region, while sibling clusters could represent other types of tumours.
  \item Agglomerative or bisecting deglomerative hierarchical algorithms provide limited information about siblings since only two objects are merged or divided on each level. Therefore, many siblings are mistreated as an artificial parent-child relation. Such mistreatment requires an additional post-processing step.
\end{itemize}

\begin{figure}[h!]
  \includegraphics[width=0.45\textwidth]{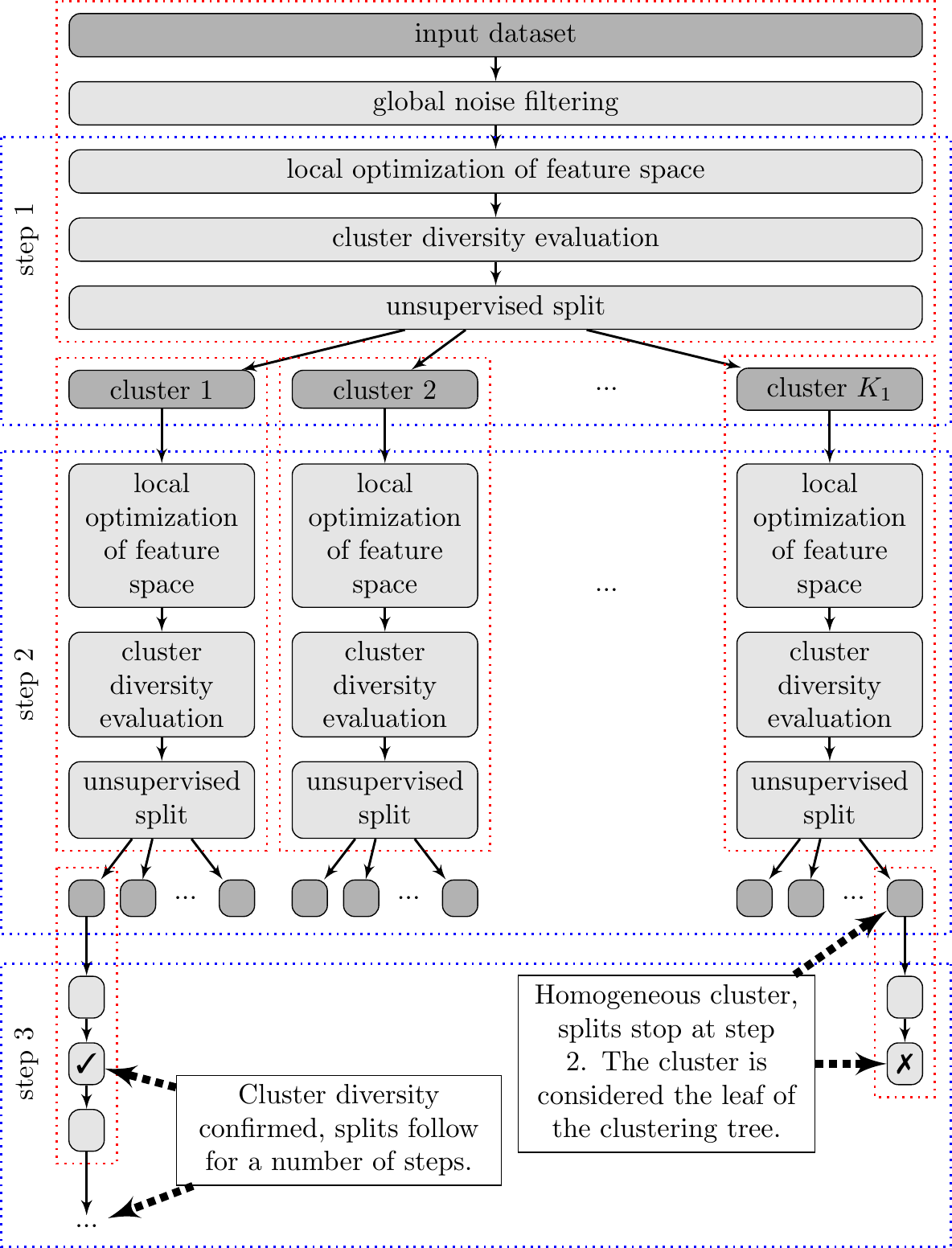}
  \centering
  \caption{The proposed hierarchical framework for unsupervised clustering of MSI data with local feature domain adaptation. For each ROI, an analysis was performed on whether it contains any implicit structures. If structures were present, molecular segmentation was continued recursively for each cluster/subcluster.}
  \label{fig:divik_chart}
\end{figure}

We propose a hybrid framework that directly answers the above drawbacks: Divisive intelligent K-Means (DiviK). It is a stepwise clustering procedure with feature engineering applied locally at each step (see Figure \ref{fig:divik_chart}). We implemented local feature engineering as filtering based on GMM decomposition of the feature variance across the subregion. The clustering algorithm implemented in our approach is a standard K-Means algorithm with a tailored procedure for initial condition selection. One can easily substitute GMM filtering and K-Means with other algorithms. Information criteria decide the number of clusters in the optimised feature space. The decision on the cluster number is two-stage: due to its numerical complexity, we use GAP at the first stage when assessing the data's homogeneity. We use the approximate Dunn's index to select the best local partition in the next step.

The strength of the proposed solution lies in optimising the feature space locally for data clustering according to the variability of the analysed subregion. DiviK considers separate regions independently on each step, starting from the entire input dataset feature space. It provides biologically relevant results and sustains scalability for large datasets, contrary to sophisticated approaches.

DiviK addresses the drawbacks mentioned above in the following way:

\begin{itemize}
  \item It has a hierarchical nature to provide the most comprehensive insight.
  \item During the analysis, local domain adaptation is performed separately for each cluster/subcluster to avoid the excessive influence of dominating global patterns.
  \item Before the cluster/subcluster analysis, a check is performed on whether there is evidence that diverse groups are present in the cluster/subcluster.
\end{itemize}

\section*{Implementation}

Three main components of the DiviK algorithm: feature engineering and filtering technique, stop condition definition, and clustering algorithm, constitute its framework. All of these can be easily changed to adapt to the specificity of the analysed data.

\subsection*{Feature Space Optimisation}

\begin{figure}[h!]
  \includegraphics[width=0.45\textwidth]{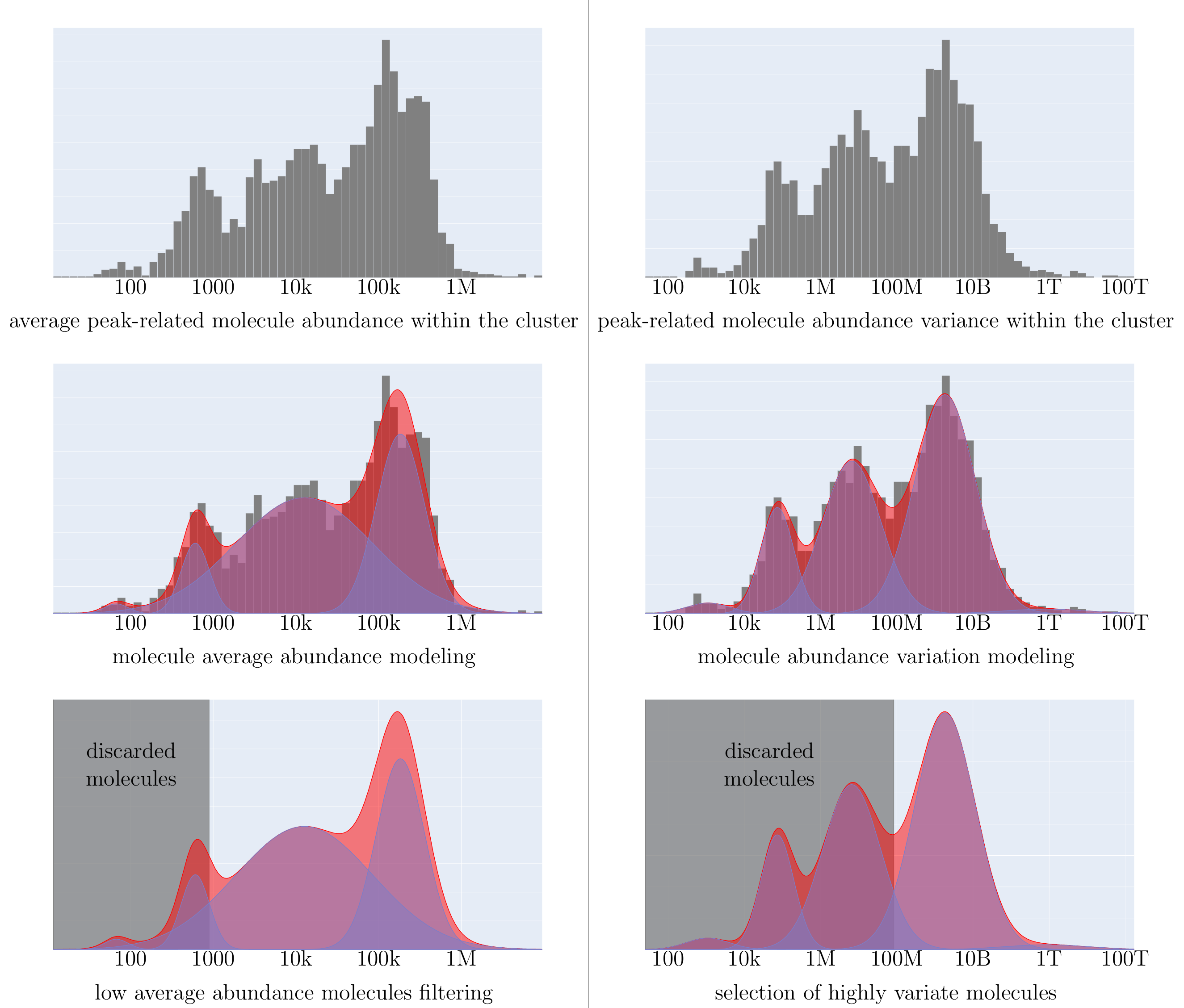}
  \centering
  \caption{Data-driven MSI peaks selection based on a histogram decomposition. Filtering considers each MSI peak's average abundance (left panel) and abundance variance (right panel). The histograms of these two characteristics are decomposed into a GMM. Components represent sets of features similar concerning selected characteristics, i.e., average abundance or abundance variance across a cluster. We calculate a conditional probability for each Gaussian component for each value of the selected characteristic. Then we apply the maximum classification rule, which leads to the interpretation that the crossing points of the neighbouring GMM components become filtering thresholds. We remove all the peaks represented by the first GMM component for the average abundance. For the abundance variance, we persist only the peaks represented by the topmost GMM components, but not less than 1\% of all the peaks.}
  \label{fig:feature_characteristic_decomposition}
\end{figure}

We propose to use a feature filtering procedure based on GMM decomposition \cite{marczyk2013adaptive,polanski2018initializing}. First, we average abundance and calculate abundance variance across the analysed subcluster for each peak. Then, empirical distributions of average abundances and abundance variances across all peaks, presented on the logarithmic scale (upper panel of Figure \ref{fig:feature_characteristic_decomposition}), are decomposed into Gaussian mixtures, as shown in the middle panel of Figure \ref{fig:feature_characteristic_decomposition}. We use the maximum conditional probability criterion to classify a particular peak into one of the mixture components. The crossing points between Gaussian components serve as the thresholds.

We discarded the GMM components containing less than 1\% of peaks to avoid numeric artefacts.

We filter out all the peaks with the average abundance below the first crossing point of the GM model. For the abundance variance, we persisted only the peaks with the variance above the topmost crossing point, assuming it results in no less than 1\% of the available features, as is shown in the bottom right panel of Figure \ref{fig:feature_characteristic_decomposition}. If the 1\% condition is not fulfilled, the second left or the second topmost threshold is used.

As we aim for a hands-free pipeline, Bayesian Information Criterion is used to select the optimal number of GMM components for both the model of average abundance distribution for global noise filtering and the model of local abundance variance distribution for discrimination of locally the most informative features.

\subsection*{Stop Condition}

\begin{figure}[h!]
  \includegraphics[width=0.45\textwidth]{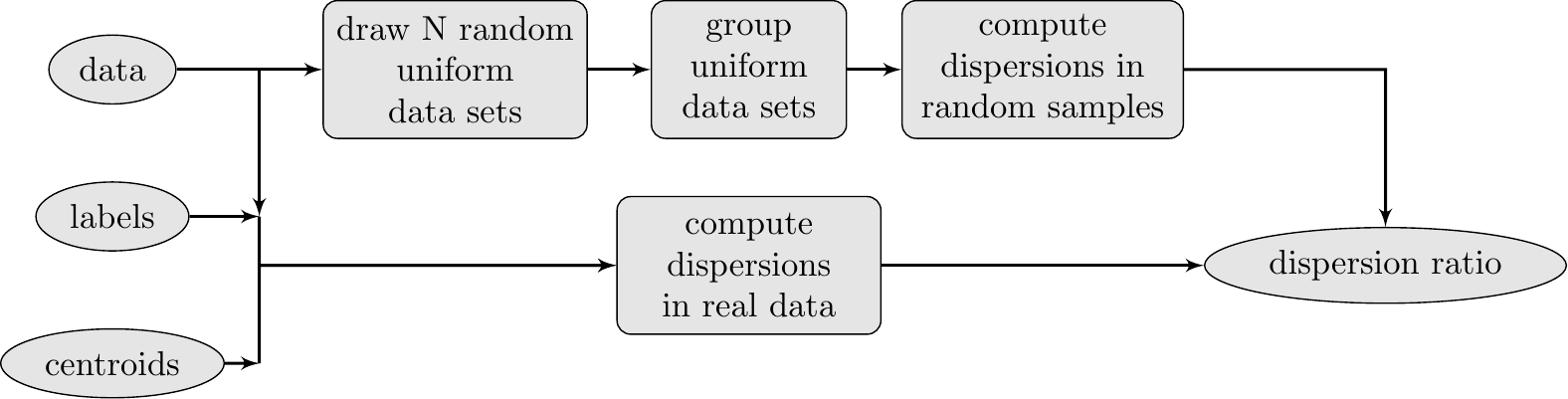}
  \centering
  \caption{Flowchart of GAP statistic computation used in the two-trial scenario. Spectra belonging to an ROI are artificially clustered into a single cluster and then into two clusters. The GAP statistic was compared for both cases. If spectra form a single group, the GAP statistic tends to be greater for the artificial division. Otherwise, it tends to be greater for two clusters (but not necessarily optimal).}
  \label{fig:gap_chart}
\end{figure}

Having a feature space optimised for cluster discovery, we must assess whether there are any clusters. There exist many indices that can validate the quality of unsupervised segmentation. These take into account cluster separability, compactness or probabilistic measures. However, only a few allow for a comparison between a multi-cluster partition and no further partition (one cluster only). For the K-Means algorithm, the GAP statistic \cite{tibshirani2001estimating} is one of the options. It relates a data-specific partition to partitions obtained via the same method over random datasets within the same bounds (see Figure \ref{fig:gap_chart}). We stop splitting of particular ROI if GAP statistic suggests region homogeneity.

\subsection*{Clustering}

In this work, we focus only on the K-Means-based implementation due to its promising initial results \cite{widlak2016detection,pietrowska2017molecular} and low computational complexity. With a cluster molecular diversity confirmed, we can proceed with the clustering. For MSI data analysis, we adjusted the distance metric \cite{widlak2016detection} and the initialisation method.

The K-Means algorithm requires the number of clusters to be specified before the analysis occurs. Since the actual number of molecularly homogeneous regions is unknown, we use unsupervised quality metrics to guide the computations. The current implementation includes Dunn's index \cite{dunn1974well} and GAP statistic \cite{tibshirani2001estimating}.

\subsubsection*{Initialisation Method}

\begin{figure}[h!]
  \includegraphics[width=0.45\textwidth]{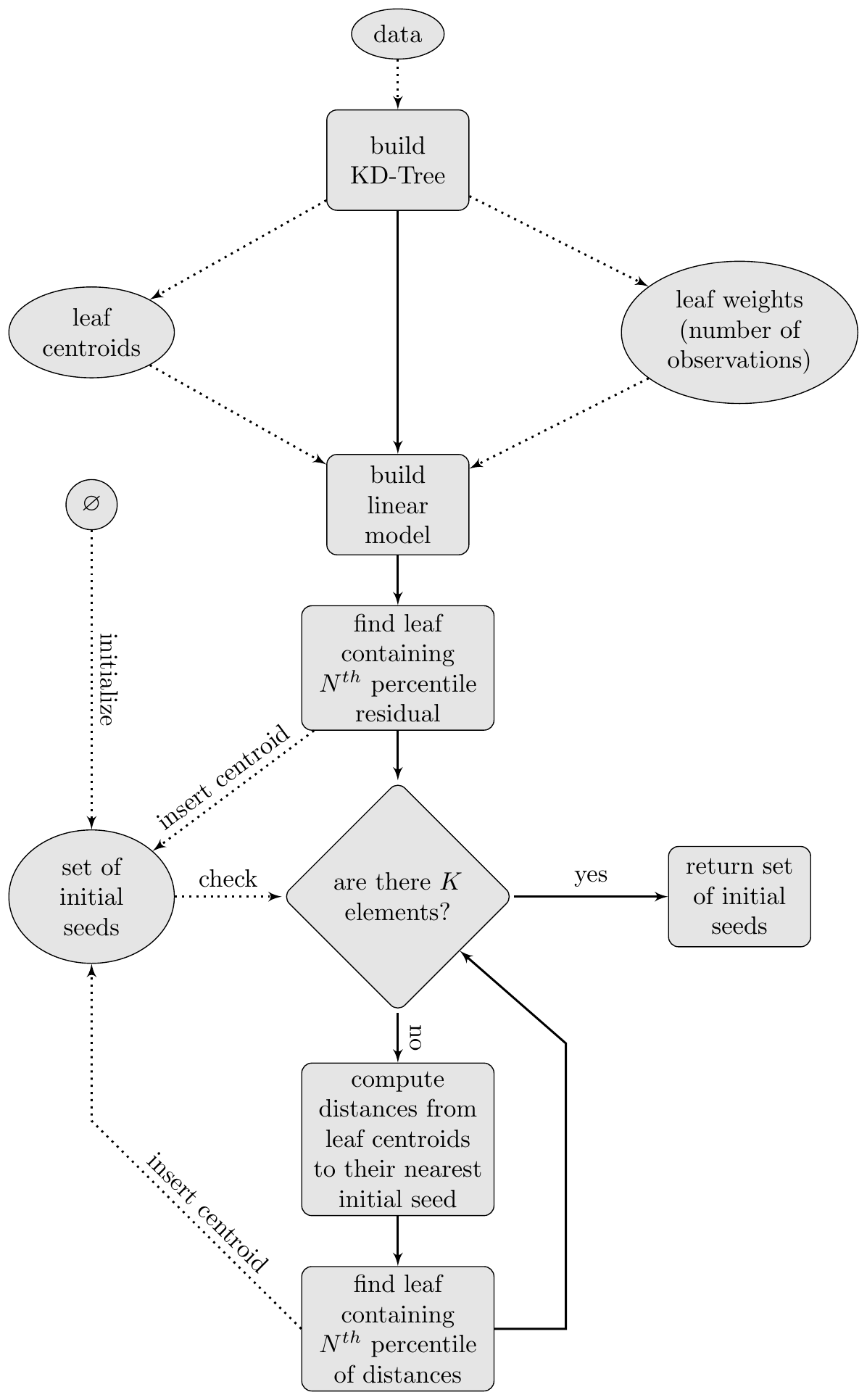}
  \centering
  \caption{Flowchart of K-Means initialization procedure. We find a moderately outlying KD-Tree leaf centroid as the first seed and follow an iterative deterministic algorithm similar to the K-Means++ algorithm. KD-Tree keeps the order of computational complexity low, while the linear model makes it more predictable than the K-Means++ algorithm.}
  \label{fig:kmeans_init_chart}
\end{figure}

Since the DiviK algorithm aims to identify biologically meaningful heterogeneity in the data, we decided to implement the deterministic initialisation procedure (see Figure \ref{fig:kmeans_init_chart}). Although there are plenty of well-known algorithms for random initialisation, the deterministic approaches are usually customised to the data structure, allowing to achieve the clustering goal. As mentioned above, in the case of MSI data, we seek the most heterogeneous tissue regions independently of the cluster imbalance. To address the expected redundancy in MSI data and decrease the computational time, we compress the data by the KD-tree algorithm (with the compression factor equal to 0.01 for 2D data and 0.001 for 3D data). The obtained "boxes of spectra" are represented by their centroids and weighting factors related to the number of spectra assigned to the particular box. The next step requires computing the distances from the global dataset centroid to the box-specific centroids. We approximate the distances' cumulative distribution function (cdf) and choose the centroid containing the $k$-th percentile. It serves as the first initial seed for the clustering algorithm. The procedure mentioned above repeats to obtain the second seed, but we substitute the global centroid with the first seed and the box containing that seed is not considered in the calculations. Having two initial seeds, we can search for the third one (if required). We estimate the cdf of the distance between the box-specific centroids and the line defined by already chosen two seeds to find the box containing $k$-th percentile. The centroid of that box is considered the third seed. One can identify the subsequent seeds similarly, substituting the line by the hyperplane defined by the already found seeds. We perform the initial seed setting in the already locally optimised feature space. The comprehensive study on the initialisation procedures (not shown in the manuscript) demonstrated the power and robustness of our approach when compared to the commonly used initialisation algorithms.

\subsection*{Scalability}

DiviK is an original stepwise deglomerative algorithm that provides scalability for big data in several ways:

\begin{itemize}
  \item It uses a locally optimised K-Means algorithm iteratively. K-Means algorithm is known as one of the few highly scalable clustering algorithms.
  \item Filtration-based feature engineering, used in the local optimisation of the feature space step, requires conducting neither PCA nor UMAP. The Gaussian Mixture Model approach supports threshold identification necessary for the feature filtration (see Figure \ref{fig:feature_characteristic_decomposition}). It is computationally efficient, as it requires only computing the within-cluster mean and the variance of each feature, followed by an automated data-driven threshold selection.
  \item To ensure computational complexity is linear with the number of samples, we developed a procedure of deterministic initialisation based on a KD-Tree algorithm.
  \item The automatic selection of the number of clusters in the K-Means algorithm is computationally expensive by default. Therefore, we split the decision process into two steps. First, we use the sampled distribution of GAP statistics to decide if the cluster is heterogeneous (requires further splitting) or homogenous (the splitting process is finished). In the case of a heterogeneous cluster, we use Dunn's Index to decide on the number of subclusters, which keeps the process scalable.
\end{itemize}

\subsection*{Technology}

DiviK was written mainly in Python, and it is distributed cross-platform under the Apache 2.0 license through Python Package Index (\url{https://pypi.org/project/divik/}) and Docker Hub (\url{https://hub.docker.com/r/gmrukwa/divik}). It is designed to be simple, efficient, accessible to a non-expert, and highly reusable. Thus, DiviK is accessible in Python directly and via a command-line interface (CLI). The Python API follows the scikit-learn \cite{sklearn2013api} design patterns and similar package organisation conventions to provide reusable building blocks. The CLI allows the construction of highly-flexible processing pipelines owing to its injection-based configuration system.

\section*{Results}

In order to compare the effectiveness of the DiviK framework and the existing approaches, we evaluated the tool against two MSI datasets of different characteristics:

\begin{itemize}
  \item Oral Squamous Cell Carcinoma (OSCC). It consists of two individual slices collected from different patients suffering from OSCC, annotated by an experienced pathologist;
  \item Mouse kidney. It consists of 75 serial slices from a single 3D-allocated volume from a mouse kidney without annotation.
\end{itemize}

We compared DiviK to the combinations of state-of-the-art MSI data clustering and global feature engineering methods. We selected K-Means, spectral and spatial clustering \cite{alexandrov2011efficient} as representatives of clustering algorithms validated for MSI. All were used in a fully unsupervised setting with the number of clusters selected using the GAP statistic. For global feature engineering, we used one of the following options: no global feature engineering, PCA with the knee-based selection of the number of components \cite{satopaa2011finding}, PCA on top of EXIMS-preselected features \cite{wijetunge2015exims}, UMAP \cite{mcinnes2018umap}, and neural ion images obtained with Xception network \cite{zhang2021spatially}.

All the experiments were carried out in the Polyaxon environment \cite{mourafiq2017polyaxon}.

\subsection*{Oral Squamous Cell Carcinoma}

\begin{figure}[h!]
  \includegraphics[width=0.45\textwidth]{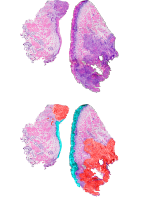}
  \centering
  \caption{Microscope image of raw HNC samples (upper panel) and samples annotated by a pathologist (lower panel). Regions of interest (ROI): red = tumour; cyan = healthy epithelium; magenta = other healthy tissue. Due to medical relevance, we focused mostly on tumour and epithelium tissue -- the origin of OSCC.}
  \label{fig:he_ground_truth}
\end{figure}

The OSCC dataset \cite{widlak2016detection} is a two-dimensional MALDI-ToF MSI dataset. The original dataset consists of 45,738 raw spectra with 109,568 mass channels. After pre-processing and spectra modelling, 3,714 peaks/features were obtained. The spectra are provided with annotations of the tissue type, allowing for distinguishing regions of interest (ROI) of the tumour, epithelium, and healthy areas (Figure \ref{fig:he_ground_truth}). Therefore, the OSCC dataset can be a benchmark for classification and clustering problems. There are diverse sources of variability, including:

\begin{itemize}
  \item inter-patient sample heterogeneity;
  \item tissue type heterogeneity;
  \item intra-tissue heterogeneity.
\end{itemize}

\begin{figure}[h!]
  \includegraphics[width=0.45\textwidth]{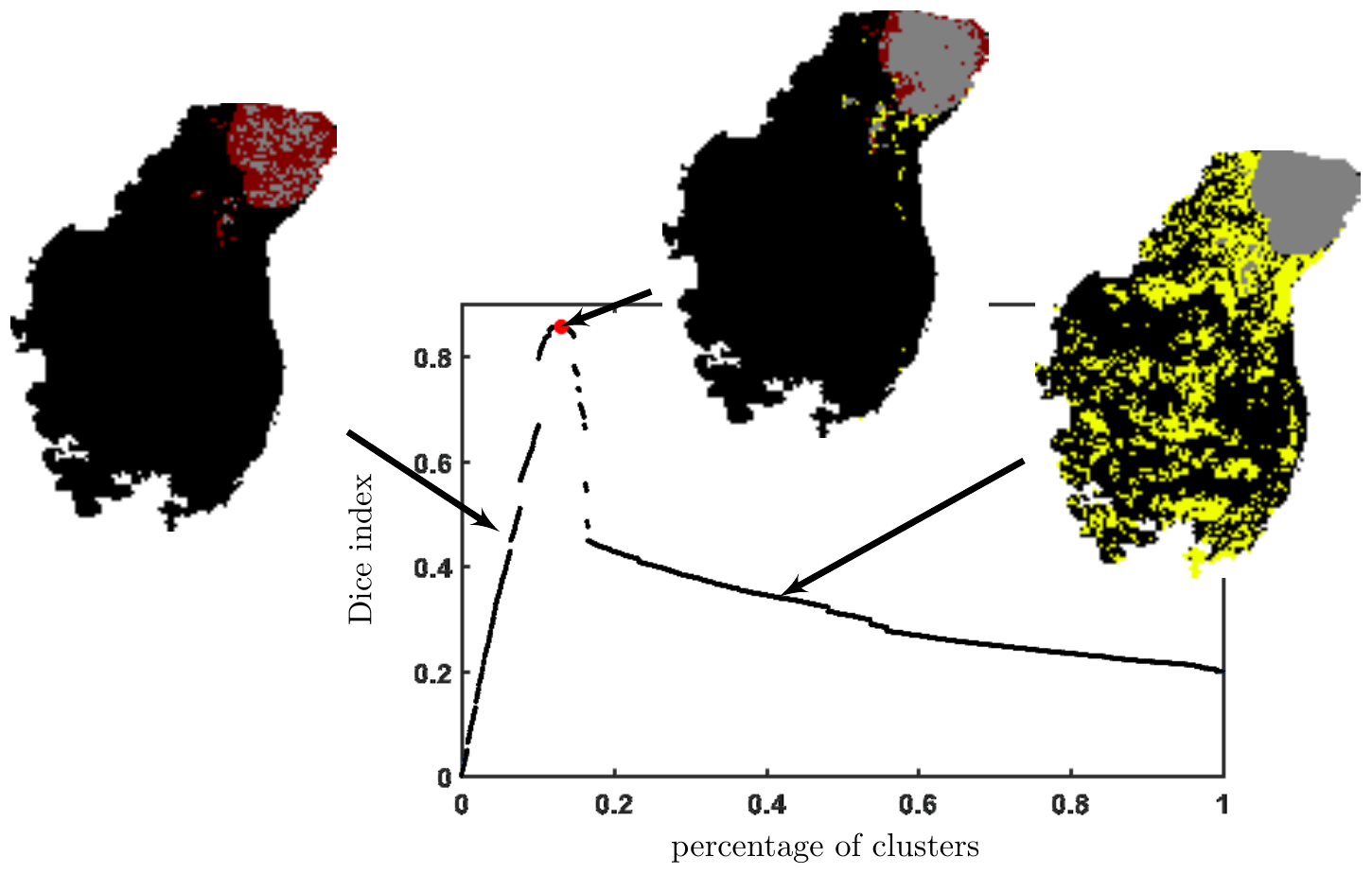}
  \centering
  \caption{Cluster label normalisation procedure. Red = false negative; yellow = false positive; grey = true positive tumour. In the first step, a binary decision was made on whether the label should belong to one of the pathologist-defined regions. Clusters were sorted by the percentage of their area covered by the ROI. They were selected sequentially to optimise the Dice index. Secondly, all ambiguous assignments were resolved via optimising the Rand Index to form the normalised labels.}
  \label{fig:normalization_chart}
\end{figure}

We selected a subset consisting of two patient samples with top tissue variety (19,874 spectra and 3,714 peaks in total) and segmented the subset using all the combinations of the methods discussed in the previous section. The obtained clusters of spectra usually consisted of spectra originally annotated to different tissue subtypes, so the annotation translation from spectrum to cluster level is required. The developed reannotation procedure maximises the multi-cluster Dice similarity index related to the primary tissue subtypes (Region of Interests ROI) as presented in Figure \ref{fig:normalization_chart}. First, we sort clusters according to the percentage of their area covered by the consecutive ROIs. Then, we select clusters sequentially to maximise the Dice index for a given ROI collectively. We assign one for each $(cluster, ROI)$ pair if the cluster is included in the set maximising Dice index for a given ROI, zero otherwise. If only one pair $(C, ROI_i)$ is assigned 1 for a given cluster $C$, the assignment is considered final. Otherwise, we evaluate each pair (assignment scenario) in terms of the Rand index.

We used this dataset to compare ROI reconstruction capabilities across different clustering algorithms and feature engineering techniques. With clusters reassigned to the pathologist-defined ROIs, we considered the following quality metrics:

\begin{itemize}
  \item Rand Index -- to capture global multi-ROI reconstruction capabilities;
  \item Dice Index -- to capture tumour reconstruction capabilities;
  \item EXIMS score -- to capture the spatial consistency of the clusters.
\end{itemize}

The EXIMS score is a measure with unbounded values, useful in comparative analyses, but its magnitude is hard to interpret. Therefore we scaled and clipped the values so that the highest multi-cluster result was $1$.

DiviK sweeps for up to 10 clusters on each segmentation level. It operates with correlation distance, as researchers have already proven it to work well for MSI data \cite{dexter2017two,smets2019evaluation}. The minimal number of local features required to preserve is 1\%. The minimal size of the cluster was set to 200 spectra. The leaf size of KD-Tree was 1\%, and the algorithm started from the leaf containing the 99th percentile of the distance. We sampled ten times 1,000 spectra each to compute Dunn's and GAP indices to keep the computational complexity of quality estimation bounded. Standalone K-Means clustering swept for up to 50 clusters, with the same criteria for computing the GAP index as DiviK and the correlation distance. Spatial clustering was launched with a radius of 7. Spectral clustering was used with cosine metrics during the embedding. The embedding generated the number of components equal to 1\% of the number of global features (information capacity equivalent to DiviK filtering). UMAP was run with 30 neighbours, correlation distance, 500 epochs and a negative sample rate of 70 to obtain three components. We used the Xception network with a patch size of 71x71 pixels and an upsample rate of 4 to ensure a patch edge of $1.775mm$.

\begin{figure}[h!]
  \includegraphics[width=0.45\textwidth]{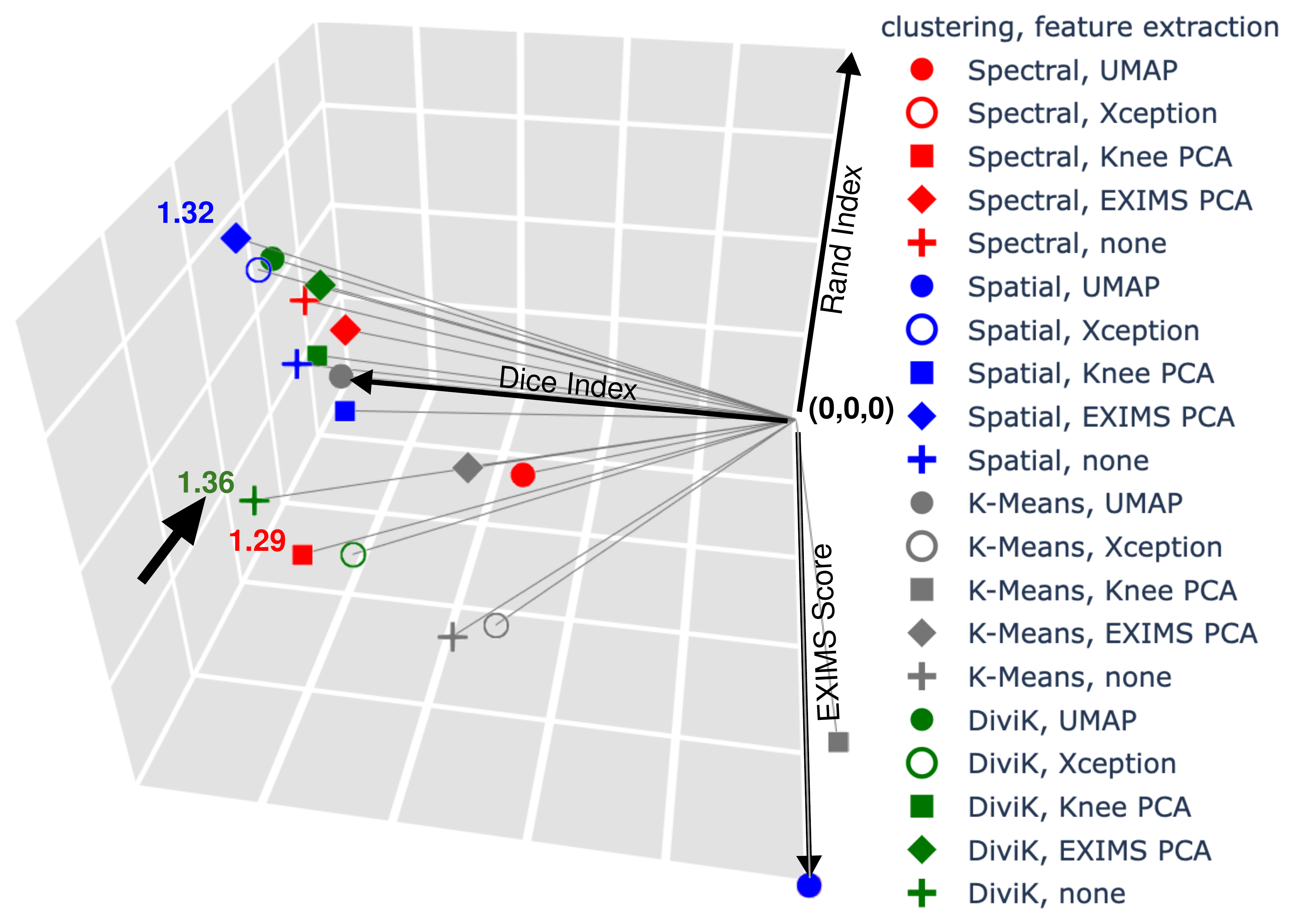}
  \centering
  \caption{Quality indices of ROI composability by the obtained clusters. The \emph{overall quality} is defined as the vector's length limited by the quality indices' values and the origin of the coordinate system. The length of the vector is displayed next to the points representing the top results. The arrow indicates the best result.}
  \label{fig:quality_indices}
\end{figure}

\begin{figure*}[h!]
  \includegraphics[width=0.95\textwidth]{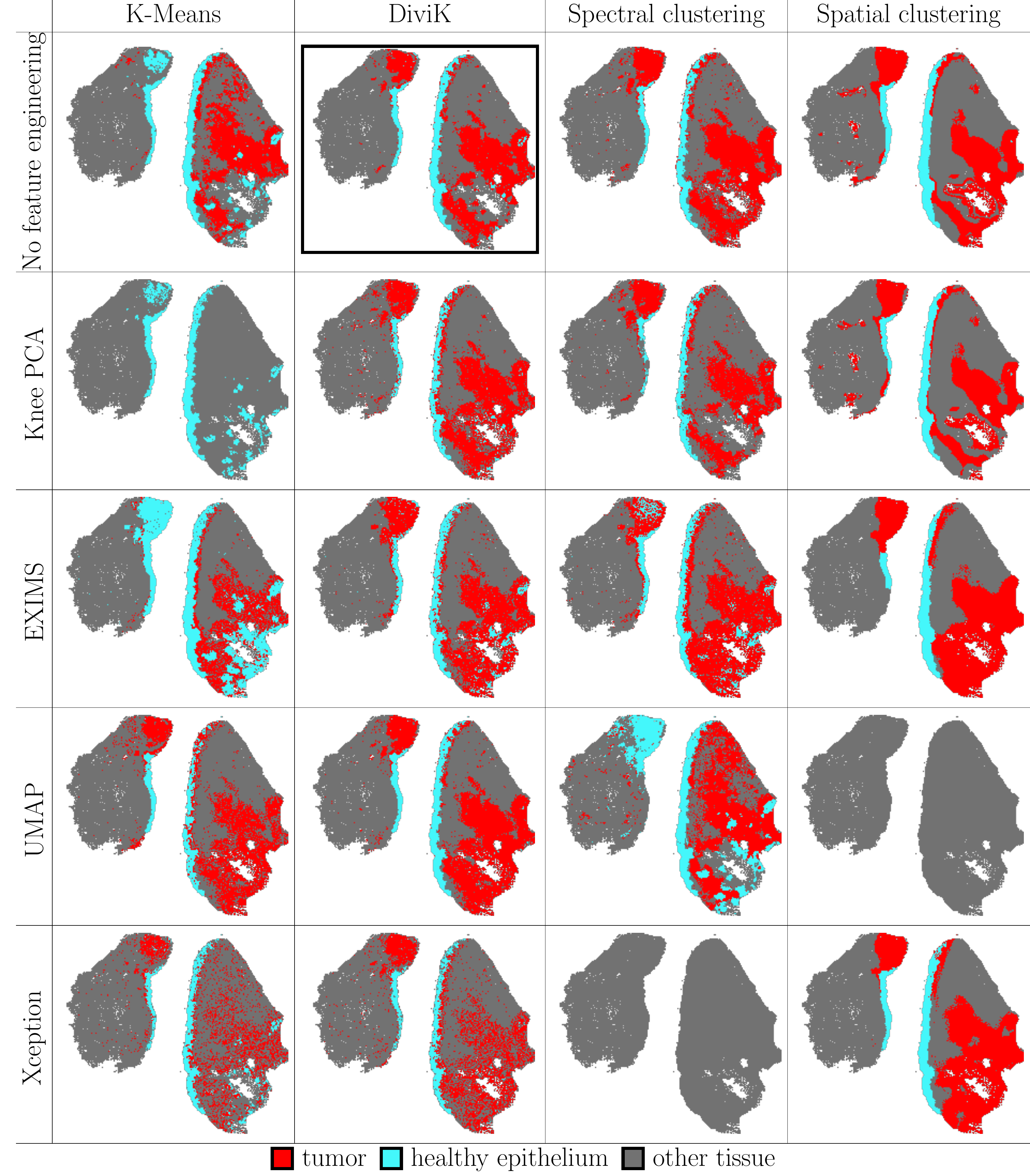}
  \centering
  \caption{Results of the unsupervised analyses for OSCC data. Clusters were matched to the pathologist regions using the method described. Red = tumour region; cyan = healthy epithelium; grey = other tissue. The top \emph{overall quality} is marked.}
  \label{fig:clusters_io}
\end{figure*}

\begin{figure}[h!]
  \includegraphics[width=0.45\textwidth]{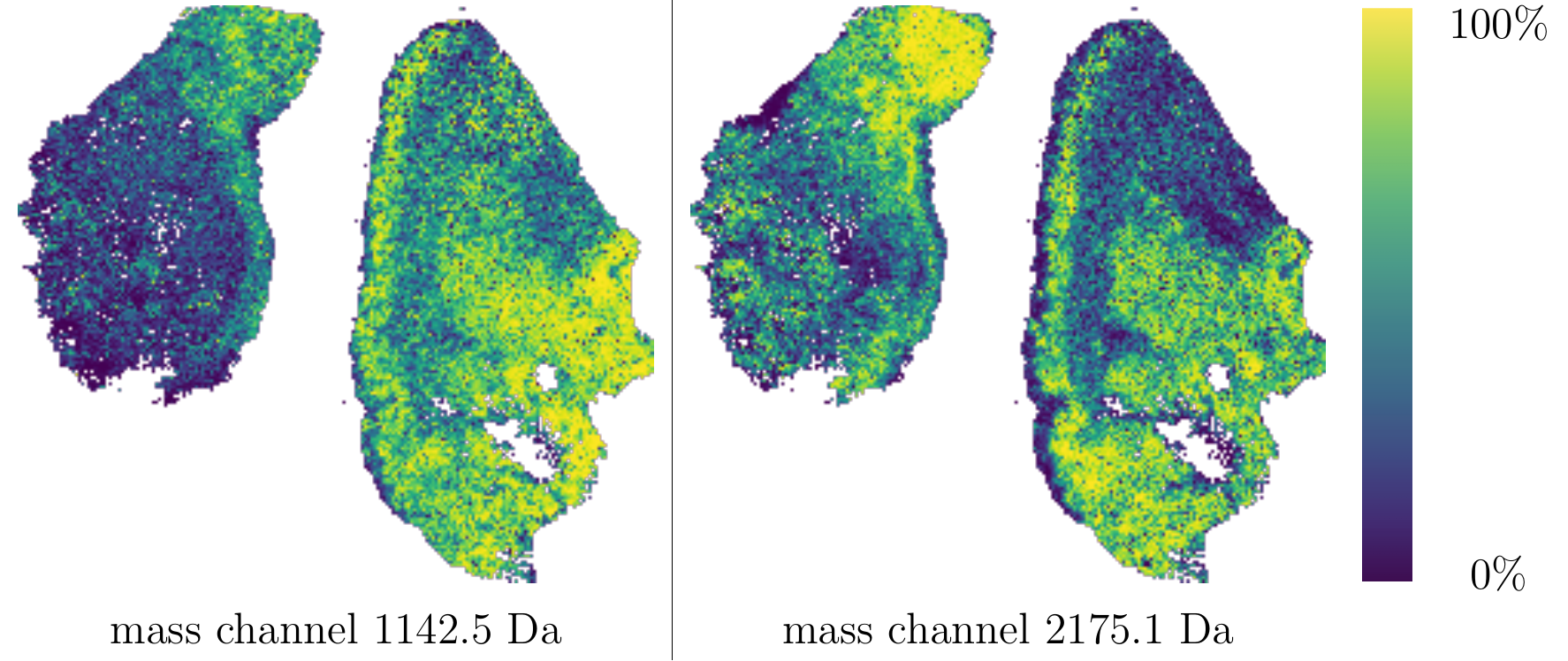}
  \centering
  \caption{The spatial distribution of the chosen features (mass channels) in OSCC data. Peptides upregulated in tumour (represented by 1142.5 m/z and 2175.1 m/z) are putatively fragments of pyruvate kinase, an enzyme involved in the Warburg effect.}
  \label{fig:clusters_io_mzs}
\end{figure}

We presented the visualisation of the performance of the DiviK algorithm and remaining combinations in Figure \ref{fig:quality_indices}. Dimensions of the presented cube are the quality indices mentioned at the beginning of this section, namely Dice Index, Rand Index, and EXIMS Score. The exact values of quality metrics are available in Table \ref{tab:quality}. In Figure \ref{fig:clusters_io} one can find a normalised map of the clusters obtained in the process. We compare the combinations of global feature engineering methods and clustering algorithms exhaustively. The proposed DiviK approach uses no global feature engineering method, yet we conducted such experiments and presented their results for comparison. Sample spatial ion heatmaps, correlated with the pathologist's tissue subtypes and the clusters found by DiviK, are presented in Figure \ref{fig:clusters_io_mzs}. The averaged results for the used clustering methods can be seen in Table \ref{tab:stability_ctr} whereas Table \ref{tab:stability_fe} presents results obtained with the used global feature engineering methods for $d$-based aggregations. Tables \ref{tab:summary} and \ref{tab:summary_fe} present the rank-based summarisation results.

\begin{table*}[h!]
  \caption{Values of quality indices measured for OSCC data. Top three values of a quality index are coloured in grey (darker is better).}
  \label{tab:quality}
  \begin{tabular}{
    m{0.07\textwidth}<{\centering\arraybackslash}
    m{0.09\textwidth}<{\centering\arraybackslash}
    |
    m{0.04\textwidth}<{\centering\arraybackslash}
    m{0.04\textwidth}<{\centering\arraybackslash}
    m{0.05\textwidth}<{\centering\arraybackslash}
    |
    m{0.04\textwidth}<{\centering\arraybackslash}
    m{0.04\textwidth}<{\centering\arraybackslash}
    m{0.05\textwidth}<{\centering\arraybackslash}
    |
    m{0.04\textwidth}<{\centering\arraybackslash}
    m{0.07\textwidth}<{\centering\arraybackslash}
    m{0.07\textwidth}<{\centering\arraybackslash}
  }
    \toprule
      \makecell{clustering\\algorithm} &
      \makecell{global\\feature\\engineering\\method} &
      \makecell{adjusted\\Rand\\index} &
      \makecell{Dice\\index} &
      \makecell{relative\\EXIMS\\score} &
      \makecell{adjusted\\Rand\\index\\rank} &
      \makecell{Dice\\index\\rank} &
      \makecell{relative\\EXIMS\\score\\rank} &
      \makecell{sum of\\ranks} &
      \makecell{overall\\quality\\$d(0,0,0)$} &
      \makecell{overall\\quality\\$d(1,1,1)$} \\
    \midrule
    Spectral & UMAP      & 0.2792          & 0.4844          & 0.5891          & 17   & 16 & 20   & 53                      & 0.8122                    & 0.9768                    \\
    Spatial  & UMAP      & 0.0000          & 0.0000          & 1.0000          & 19.5 & 19 & 2.5  & 41                      & 1.0000                    & 1.4142                    \\
    Spectral & Xception  & 0.0000          & 0.0000          & 1.0000          & 19.5 & 19 & 2.5  & 41                      & 1.0000                    & 1.4142                    \\
    K-Means  & Knee PCA  & 0.2723          & 0.0000          & 1.0000          & 18   & 19 & 2.5  & 39.5                    & 1.0364                    & 1.2368                    \\
    K-Means  & Xception  & 0.3098          & 0.4577          & 0.9197          & 16   & 17 & 8    & 41                      & 1.0730                    & 0.8814                    \\
    K-Means  & EXIMS PCA & 0.4827          & 0.5129          & 0.8323          & 12   & 14 & 9    & 35                      & 1.0903                    & 0.7300                    \\
    Spectral & EXIMS PCA & 0.5447          & 0.7418          & 0.6449          & 8    & 8  & 18   & 34                      & 1.1237                    & 0.6325                    \\
    K-Means  & none      & 0.3364          & 0.5043          & 0.9712          & 15   & 15 & 6    & 36                      & 1.1449                    & 0.8288                    \\
    K-Means  & UMAP      & 0.5231          & 0.7238          & 0.7225          & 10   & 10 & 13   & 33                      & 1.1487                    & 0.6170                    \\
    Spatial  & Knee PCA  & 0.4985          & 0.7065          & 0.7639          & 11   & 11 & 10   & 32                      & 1.1537                    & 0.6272                    \\
    DiviK    & EXIMS PCA & 0.6082          & 0.7765          & 0.6383          & 4    & 5  & 19   & 28                      & 1.1749                    & 0.5782                    \\
    Spectral & none      & 0.5906          & 0.7966          & 0.6520          & 5    & 4  & 17   & 26                      & 1.1868                    & 0.5745                    \\
    DiviK    & Knee PCA  & 0.5567          & 0.7540          & 0.7289          & 7    & 7  & 12   & 26                      & 1.1873                    & 0.5749                    \\
    DiviK    & Xception  & 0.4203          & 0.6429          & 0.9395          & 14   & 13 & 7    & 34                      & 1.2136                    & 0.6835                    \\
    Spatial  & none      & 0.5617          & 0.7720          & 0.7587          & 6    & 6  & 11   & 23                      & 1.2195                    & 0.5498                    \\
    DiviK    & UMAP      & 0.6534          & 0.8369          & 0.6568          & 2    & 3  & 16   & 21                      & 1.2485                    & \cellcolor{gray!25}0.5143 \\
    Spatial  & Xception  & 0.6517          & 0.8465          & 0.6851          & 3    & 2  & 15   & \cellcolor{gray!45}20   & 1.2691                    & \cellcolor{gray!45}0.4940 \\
    Spectral & Knee PCA  & 0.4594          & 0.6897          & 0.9891          & 13   & 12 & 5    & 30                      & \cellcolor{gray!25}1.2904 & 0.6235                    \\
    Spatial  & EXIMS PCA & 0.7035          & 0.8672          & 0.6977          & 1    & 1  & 14   & \cellcolor{gray!65}16   & \cellcolor{gray!45}1.3167 & \cellcolor{gray!65}0.4438 \\
    DiviK    & none      & 0.5433          & 0.7372          & 1.0000          & 9    & 9  & 2.5  & \cellcolor{gray!25}20.5 & \cellcolor{gray!65}1.3560 & 0.5269                    \\
    \bottomrule
  \end{tabular}
  \end{table*}
  
\begin{table}[h!]
  \caption{Average quality indices for clustering methods with OSCC data. Standard deviation is presented in brackets.}
  \label{tab:stability_ctr}
  \begin{tabular}{
      c
      |
      m{0.04\textwidth}<{\centering\arraybackslash}
      m{0.04\textwidth}<{\centering\arraybackslash}
      m{0.05\textwidth}<{\centering\arraybackslash}
      |
      m{0.05\textwidth}<{\centering\arraybackslash}
      m{0.05\textwidth}<{\centering\arraybackslash}
    }
    \toprule
      \makecell{clustering\\algorithm} &
      \makecell{adjusted\\Rand\\index} &
      \makecell{Dice\\index} &
      \makecell{relative\\EXIMS\\score} &
      \makecell{overall\\quality\\$d(0,0,0)$} &
      \makecell{overall\\quality\\$d(1,1,1)$} \\
    \midrule
    Spectral & \makecell{0.375\\(0.241)}                     & \makecell{0.543\\(0.325)}                     & \makecell{0.775\\(0.202)}                   & \makecell{1.083\\(0.184)}            & \makecell{0.844\\(0.357)} \\
    \hline
    K-Means  & \makecell{0.385\\(0.111)}                     & \makecell{0.440\\(0.266)}                     & \makecell{\textbf{0.889}\\\textbf{(0.113)}} & \makecell{1.099\\(0.048)} & \makecell{0.859\\(0.234)} \\
    \hline
    Spatial  & \makecell{0.483\\(0.281)}                     & \makecell{0.638\\(0.363)}                     & \makecell{0.781\\(0.127)}                   & \makecell{1.192\\(0.123)}            & \makecell{0.706\\(0.402)} \\
    \hline
    DiviK    & \makecell{\textbf{0.556}\\\textbf{(0.088)}} & \makecell{\textbf{0.750}\\\textbf{(0.071)}} & \makecell{0.793\\(0.167)}                   & \makecell{\textbf{1.236}\\\textbf{(0.073)}} & \makecell{\textbf{0.576}\\\textbf{(0.067)}} \\
    \bottomrule
  \end{tabular}
\end{table}

\begin{table}[h!]
  \caption{Average quality indices for global feature engineering methods with OSCC data. Standard deviation is presented in brackets.}
  \label{tab:stability_fe}
  \begin{tabular}{
    c
    |
    m{0.04\textwidth}<{\centering\arraybackslash}
    m{0.04\textwidth}<{\centering\arraybackslash}
    m{0.05\textwidth}<{\centering\arraybackslash}
    |
    m{0.05\textwidth}<{\centering\arraybackslash}
    m{0.05\textwidth}<{\centering\arraybackslash}
  }
    \toprule
      \makecell{global\\feature\\engineering\\method} &
      \makecell{adjusted\\Rand\\index} &
      \makecell{Dice\\index} &
      \makecell{relative\\EXIMS\\score} &
      \makecell{overall\\quality\\$d(0,0,0)$} &
      \makecell{overall\\quality\\$d(1,1,1)$} \\
    \midrule
    UMAP      & \makecell{0.364\\(0.288)}                     & \makecell{0.511\\(0.371)}          & \makecell{0.742\\(0.180)}            & \makecell{1.052\\(0.190)}                     & \makecell{0.881\\(0.407)} \\
    \hline
    Xception  & \makecell{0.346\\(0.271)}                     & \makecell{0.487\\(0.361)}          & \makecell{\textbf{0.886}\\\textbf{(0.138)}}   & \makecell{1.139\\(0.124)}                     & \makecell{0.868\\(0.397)} \\
    \hline
    Knee PCA  & \makecell{0.447\\(0.123)}                     & \makecell{0.538\\(0.359)}          & \makecell{0.871\\(0.144)}            & \makecell{1.167\\(0.105)}                     & \makecell{0.766\\(0.315)} \\
    \hline
    EXIMS PCA & \makecell{\textbf{0.585}\\\textbf{(0.094)}} & \makecell{\textbf{0.725}\\\textbf{(0.151)}} & \makecell{0.703\\(0.090)} & \makecell{1.176\\(0.100)}                     & \makecell{\textbf{0.596}\\\textbf{(0.119)}} \\
    \hline
    none      & \makecell{0.508\\(0.116)}                     & \makecell{0.703\\(0.134)} & \makecell{0.846\\(0.168)}            & \makecell{\textbf{1.227}\\\textbf{(0.091)}} & \makecell{0.620\\(0.141)} \\
    \bottomrule
  \end{tabular}
\end{table}

\begin{table}[h!]
  \caption{High-level summary of clustering algorithm characteristics. Contains sum of ranks for each criterion (lower is better).}
  \label{tab:summary}
  \begin{tabular}{ccccc}
  \toprule
                                 & \makecell{Spectral\\clustering} & K-Means       & \makecell{Spatial\\clustering} & DiviK          \\
  \midrule
  \makecell{Rand\\index\\ranks}  & 62.5                            & 71            & 40.5                           & \textbf{36}    \\
  \cmidrule(r){1-1}
  \makecell{Dice\\index\\ranks}  & 59                              & 75            & 39                             & \textbf{37}    \\
  \cmidrule(r){1-1}
  \makecell{EXIMS\\score\\ranks} & 62.5                            & \textbf{38.5} & 52.5                           & 56.5           \\
  \cmidrule(r){1-1}
  \makecell{scalability\\ranks}  & 72.5                            & \textbf{32.5} & 72.5                           & \textbf{32.5}  \\
  \midrule
  \makecell{total\\rank}         & 256.5                           & 217           & 204.5                          & \textbf{162}   \\
  \bottomrule
  \end{tabular}
\end{table}

\begin{table*}[h!]
  \caption{High-level summary of feature engineering characteristics. Contains sum of ranks for each criterion (lower is better).}
  \label{tab:summary_fe}
  \begin{tabular}{cccccc}
  \toprule
                                 & UMAP        & Xception       & Knee PCA      & EXIMS PCA   & none          \\
  \midrule
  \makecell{Rand\\index\\ranks}  & 48.5        & 52.5           & 49            & \textbf{25} & 35            \\
  \cmidrule(r){1-1}
  \makecell{Dice\\index\\ranks}  & 48          & 51             & 49            & \textbf{28} & 34            \\
  \cmidrule(r){1-1}
  \makecell{EXIMS\\score\\ranks} & 51.5        & 32.5           & \textbf{29.5} & 60          & 36.5          \\
  \cmidrule(r){1-1}
  \makecell{scalability\\ranks}  & \textbf{38} & \textbf{58}    & \textbf{38}   & \textbf{38} & \textbf{38}   \\
  \midrule
  \makecell{total\\rank}         & 186         & 194            & 165.5         & 151         & \textbf{143.5} \\
  \bottomrule
  \end{tabular}
\end{table*}

\subsection*{Mouse Kidney 3D}

The mouse kidney dataset \cite{oetjen2015benchmark} is a three-dimensional MALDI-ToF MSI dataset. Therefore, the dataset has a high volume and can be a benchmark for the algorithm's scalability. The sources of variability are limited to a single source specimen. It consists of 75 sections from the central part of a mouse kidney, 1,362,830 spectra in total, and 7,680 data points per spectrum. The dataset was already subject to Gaussian spectral smoothing and baseline reduction with the Top Hat algorithm.

We conduct clustering in the same scenarios as previously, except feature engineering based on neural ion images, as it lacks generalisation for 3D data. No expert-defined labels are available for the spectra; thus, one must assess the capabilities of heterogeneity detection visually. Clusters are subject to a similar reannotation procedure as depicted in Figure \ref{fig:normalization_chart} to indicate the similarities between results.

The minimal number of local features that DiviK is required to preserve is set to 0.5\%. We did not split the cluster further if it already was 50,000 spectra or fewer, as this dataset cannot confirm detailed internal tissue diversity. KD-Tree leaf size during the initialisation was 0.1\%, and the algorithm started from the leaf containing the 95th percentile of the distance. To compute the Dunn's and GAP indices, we sampled ten times 5,000 spectra each to keep the computational complexity of quality estimation bounded.

\begin{figure}[h!]
  \includegraphics[width=0.45\textwidth]{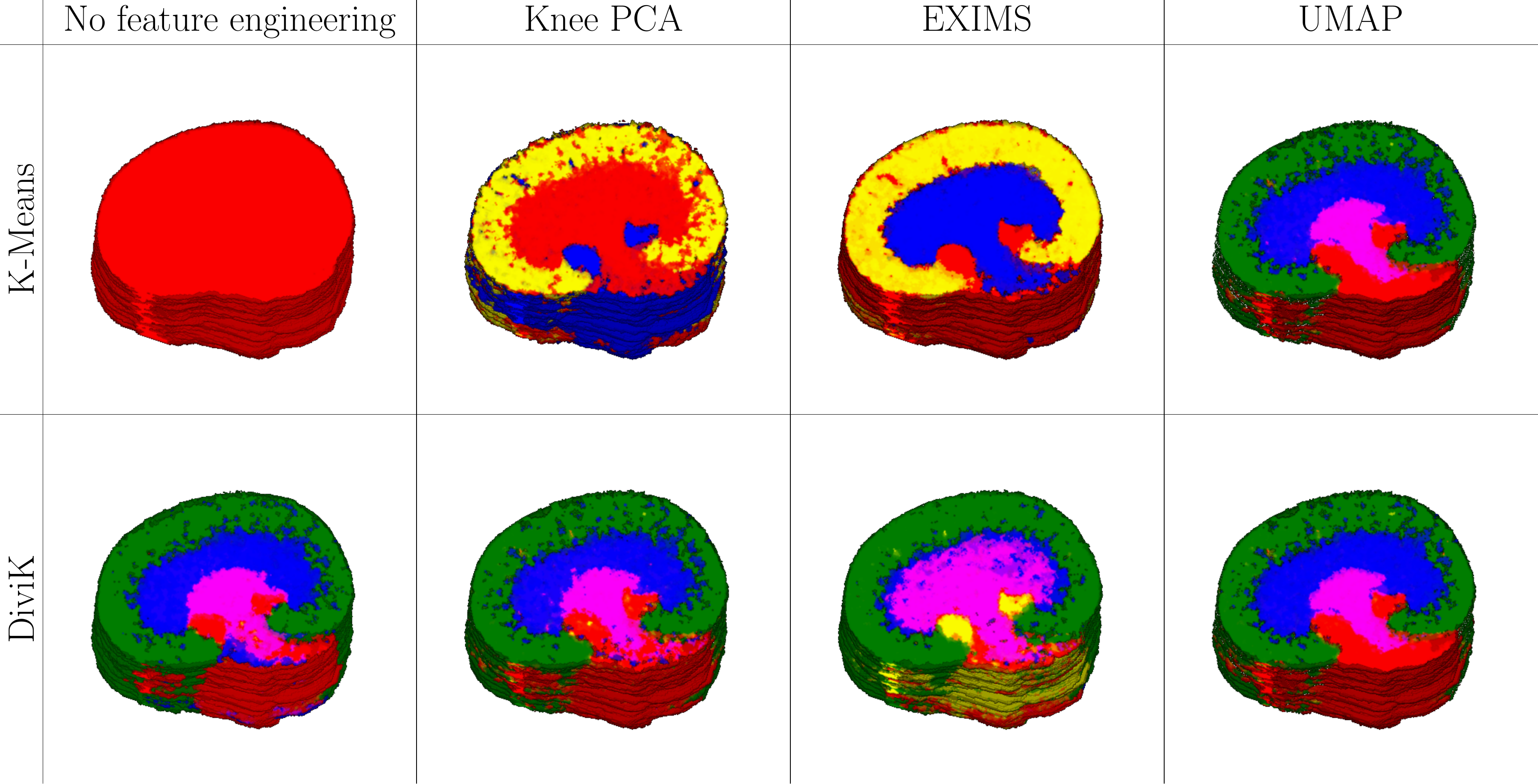}
  \centering
  \caption{Results of the unsupervised analyses for 3D mouse kidney data. All 1,362,830 spectra (75 slices) were analysed together with locally adjusted feature space. Different colours represent molecularly different tissue subtypes. We normalised the colours to correspond between experiment scenarios.}
  \label{fig:clusters_3d}
\end{figure}

\begin{figure*}[h!]
  \includegraphics[width=0.95\textwidth]{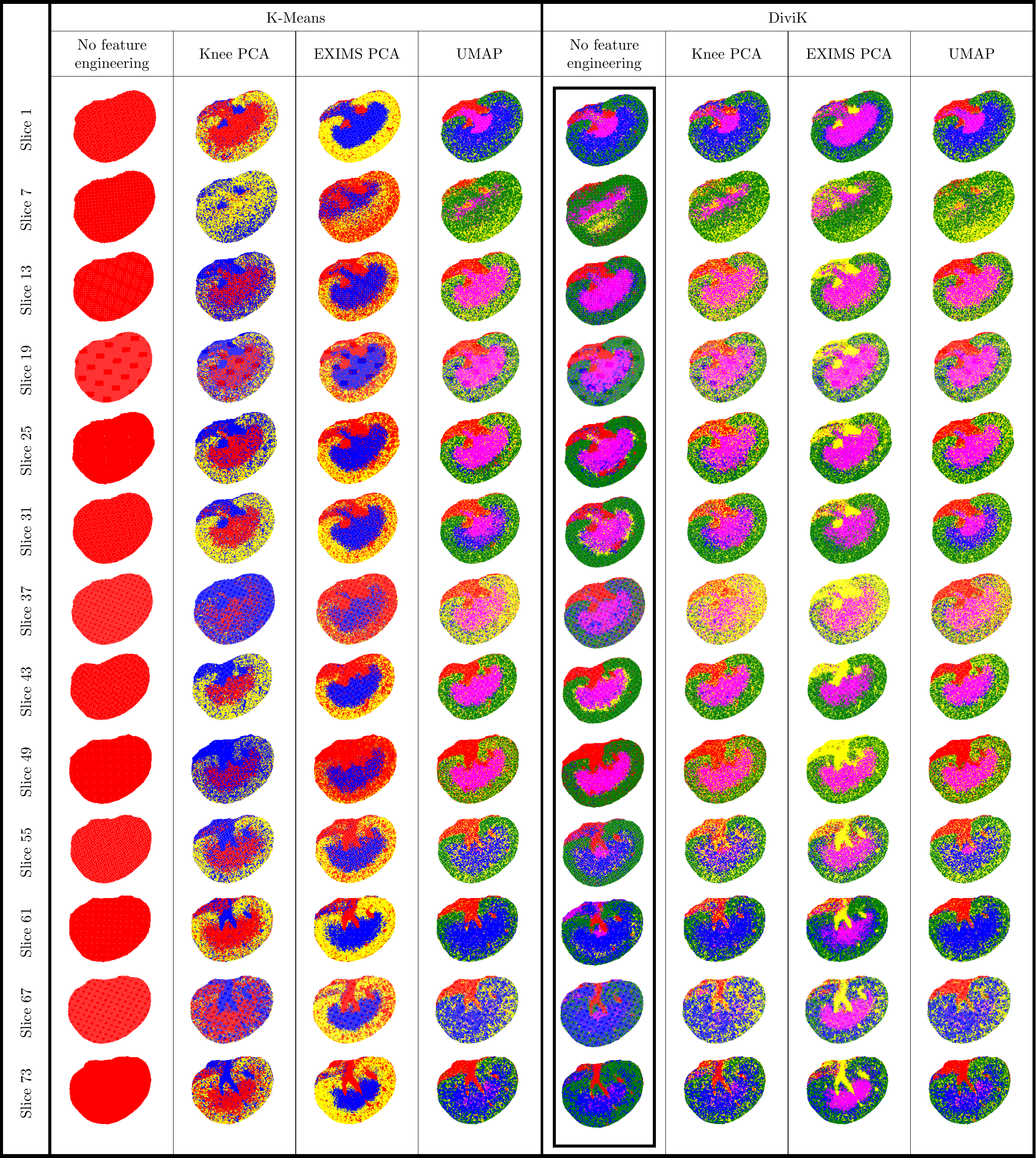}
  \centering
  \caption{Serial slices of the 3D mouse kidney data. Clusters are visualised on each sixth slice. Cluster colours were normalised to match between configurations for easy comparison.}
  \label{fig:clusters_3d_slices}
\end{figure*}

In Figure \ref{fig:clusters_3d}, we present volumes with clusters marked. We matched cluster colours to enable visual comparison, although we used no cluster normalisation procedure, as the ground truth labels are nonexistent. Unfortunately, spectral clustering is not included in this study, as the computational complexity of the basic approach did not allow for enough scalability. On the other hand, a two-step approach \cite{dexter2017two} did not lead to convergence. Additionally, spatial clustering \cite{alexandrov2011efficient} does not consider modified spatial strides in the third dimension. The same clusters are presented in Figure \ref{fig:clusters_3d_slices} on each 6th consecutive slice.

\section*{Discussion}

\subsection*{Oral Squamous Cell Carcinoma}

As shown in Figure \ref{fig:clusters_io}, it is possible to observe varying capabilities for heterogeneity detection depending on the selected methods for clustering and global feature selection. For example, UMAP combined with spatial clustering did not create clusters overlapping with biological structures (Dice index $0.00$, Rand index $0.00$, see Table \ref{tab:quality}), even though both spatial clustering and UMAP tend to exhibit the potential for capturing the necessary detail (see Figure \ref{fig:clusters_io}). One can observe similar results for the Xception-based feature engineering combined with spectral clustering.

In contrast, spatial clustering with EXIMS-based global feature engineering approximated the tumour region with the highest Dice index and the top ROIs composition expressed via Rand index. Neural ion images with spatial clustering yield the second-highest Dice index and the third-highest Rand index. However, these are not the only criteria used by medical experts during the quality assessment.

A visual comparison between clusters (Figure \ref{fig:clusters_io}) and pathologist-annotated ROIs (Figure \ref{fig:he_ground_truth}) shows that DiviK yields relatively stable regions regardless of the global feature engineering method. Table \ref{tab:stability_ctr} seems to support that statement, as the standard deviation of the Dice index and adjusted Rand index is the lowest of all clustering methods. Spectral clustering with the most basic features (no global feature engineering or knee PCA) seems to discover similar regions.

Xception-based feature engineering method resulted in much less consistent regions discovered with K-Means and DiviK. Two hypotheses seem probable explanations for this phenomenon:

\begin{itemize}
  \item Neural ion images amplified noise. Grouping of ion images deduplicates features; however, it increases the proportional representation of the noise-related mass channels compared to the original dataset. The results of spatial clustering seem to support this hypothesis, as its embedded bilateral filter reduces noise.
  \item The original ion image normalisation based on winsorisation and scaling is not sufficient for this dataset. The contrast of the patches analysed by the Xception network is too small to generate useful embedding. Since the approach proposed by the authors \cite{zhang2021spatially} formalises the visual similarity of ion maps, it could probably be useful to apply classical ion image normalisation \cite{race2015optimisation} to enhance the contrast first.
\end{itemize}

The comparison of global feature engineering methods is much less conclusive than the comparison of clustering methods. There does not seem to be any simple rule that could yield outstanding stability or performance against varying clustering methods. While EXIMS PCA and no global feature engineering seem to demonstrate increased tumour reconstruction capabilities (as expressed through the Dice index), at the same time, these yield a medium EXIMS score.

Finally, the EXIMS score and the \emph{overall quality} concept allow us to assess the trade-off between the agreement measures and the visual consistency of clusters. We defined the \emph{overall quality} in two ways:
\begin{itemize}
    \item \emph{overall quality $d(0,0,0)$} -- it is the length of the vector in 3D space defined by the values of quality indices and the origin of the coordinate system (see Figure \ref{fig:quality_indices}). The better the clustering result, the higher the value.
    \item \emph{overall quality $d(1,1,1)$} -- it is the length of the vector in 3D space defined by the values of quality indices and their maximal theoretical value in point $(1,1,1)$. The better the clustering result, the lower the value.
\end{itemize}

Naturally, the most consistent segmentations achieve the top EXIMS score: cases that yield a single cluster (spatial clustering with UMAP) or completely miss one of the ROIs (K-Means with PCA). Since these are not relevant from the medical point of view, we bind their relative EXIMS score at $1.0$ for further consideration. The DiviK clustering without a global feature engineering technique achieved the next top EXIMS score. That renders its \emph{overall quality $d(0,0,0)$} the top one. We obtained the following two best results in terms of \emph{overall quality $d(0,0,0)$} using Spatial clustering with EXIMS PCA-based feature engineering and Spectral clustering with PCA. At the same time, the \emph{overall quality $d(1,1,1)$} indicates Spatial clustering with EXIMS PCA-based feature engineering as the top result. Spatial clustering over Xception-generated features and DiviK clustering with UMAP feature engineering yield the next two best scores.

Table \ref{tab:quality} ranks the individual configurations' scores (tumour coverage, overall composition and cluster consistency) for a qualitative comparison. The lower the rank, the better the result. The lowest ranks appear consecutively for Spatial clustering with EXIMS PCA-based feature engineering, Spatial clustering over neural ion images, and DiviK with no feature engineering. Table \ref{tab:summary} continues the ranking for clustering methods. DiviK exhibits the lowest ranks for tumour coverage and overall composition, while K-Means exhibits the lowest for cluster consistency.

One can find DiviK with no feature engineering method scores within the top three in each of the aggregated overall quality measures. Sums of ranks ranged from 16 to 53. DiviK with no feature engineering was assigned a sum of ranks equal to 20.5, which is around 12\% of the observed quality measure range. Overall quality $d(0,0,0)$ ranged from 0.8122 to 1.3560, with a top value of 1.3560 for DiviK. Overall quality $d(1,1,1)$ ranged from 0.4438 to 1.4142. DiviK with UMAP feature engineering was around 7\% of the observed quality measure range.

Similarly as for the basic quality measures, Table \ref{tab:stability_ctr} and Table \ref{tab:stability_fe} summarize also both variants of \emph{overall quality}. On average, DiviK appears to be the most distant from the origin (mean $d(0,0,0)$=1.236) and the closest to the theoretical maximum of quality measures (mean $d(1,1,1)$=0.576) as its good clustering results do not depend so much on the feature engineering techniques applied. The local subcluster-driven feature space adaptation allows reducing the domain significantly by filtering the less informative and noise-level signals. The second top is the spatial clustering with mean $d(0,0,0)$=1.192 and mean $d(1,1,1)$=0.706. The Cohen's effect size in differences between these two approaches is close to medium and equals to 0.435 for $d(0,0,0)$ and 0.451 for $d(1,1,1)$. While the feature engineering techniques for other than DiviK methods are a concern, the first choice should be EXIMS PCA as it outperforms the other techniques tested. In the case of DiviK, no feature engineering is recommended.  

We conducted a qualitative evaluation of the effect of the feature engineering method and clustering algorithm on the clustering results quality. Table \ref{tab:effect_size} shows the partial $\eta^2$ effect size and Kendall's W concordance coefficient on the considered quality indices. The partial $\eta^2$ effect size values exceed the $0.14$ large effect size threshold. The choice of the clustering method influences the Dice and Rand indices more than the feature engineering method. The remaining quality measure (EXIMS score) is, as expected, stronger influenced by the feature engineering method, and the clustering method is less relevant. While an overall quality is considered, choice of clustering algorithm is more important than feature engineering techniques applied. 

\begin{table}[h!]
  \caption{Qualitive evaluation of the effect of the feature engineering method and clustering algorithm. We conducted Kruskal-Wallis test and effect size used in the Table is partial $\eta^2$. Furthermore, we calculated Kendall's W concordance index.}
  \label{tab:effect_size}
\begin{tabular}{ccc}
  \toprule
  \makecell{Quality\\measure} & \makecell{Feature\\engineering} & \makecell{Clustering} \\
  \midrule
  \multicolumn{3}{c}{Partial $\eta^2$} \\
  \hline
  Rand index & 0.203 & 0.258 \\
  Dice index & 0.161 & 0.293 \\
  EXIMS & 0.262 & 0.095 \\
  Overall quality $d(0,0,0)$ & 0.141 & 0.345 \\
  Overall quality $d(1,1,1)$ & 0.122 & 0.309 \\
  \hline \\
  \multicolumn{3}{c}{Kendall's W Concordance Index} \\
  \hline
  Rand index & 0.328 & 0.325 \\
  Dice index & 0.328 & 0.325 \\
  EXIMS & 0.136 & 0.375 \\
  Overall quality $d(0,0,0)$ & 0.424 & 0.138 \\
  Overall quality $d(1,1,1)$ & 0.472 & 0.138 \\
  \bottomrule
\end{tabular}
\end{table}

\subsection*{Mouse Kidney 3D}

Published benchmark datasets \cite{oetjen2015benchmark} are oriented explicitly on high-scale computations. The risk of missing the crucial details increases significantly for such a scale. Moreover, it renders many methods useless due to their computational complexity. Therefore, additional modifications are sometimes required \cite{dexter2017two}, yet insufficient.

As shown in Figure \ref{fig:clusters_3d}, the high-level composition of the medical sample is generally consistent between the configurations. The most straightforward configuration of K-Means clustering with no global feature engineering yields a single cluster for the whole dataset.

The structures discovered via K-Means and DiviK algorithms look consistent compared to other work \cite{abdelmoula2018interactive, race2016spectralanalysis, abdelmoula2021peak} that has analysed this data. The detected regions share a molecular signature across the 3D volume and exhibit functional differences.

Algorithms were ranked concerning their scalability limits; Table \ref{tab:summary} sums the ranks. K-Means and DiviK were the only algorithms that completed computations and were assigned the best ranks.

DiviK approach does not require applying PCA or UMAP globally. The GMM-based feature filtration/selection is applied to optimise the feature space locally. Thanks to that approach, the internal structure, dominated by the main clusters, might be revealed in the subsequent divisive steps. We used PCA and UMAP-based pipelines for reference in the comparison study, and they are not a part of the proposed DiviK algorithm. We completed these pipelines using a machine equipped with 48 CPU cores and 256 GB RAM, correspondingly in 23.8 minutes (PCA) and 198.5 minutes (UMAP).

On the contrary, the GMM-based filtering procedure for the full mouse kidney dataset completes on average within 8.66 seconds (156 ms standard deviation). Due to the deglomerative pipeline of the DiviK algorithm and local optimisation of the feature space, the filtering procedure is repeated per each splitting step, which led in the case of the mouse kidney dataset up to 60 decompositions for different subregions of this tissue sample. It gives, on average, less than 9 minutes (8.66) in total.

\section*{Conclusions}

The comprehensive comparative study considering several aspects of data exploration demonstrated that DiviK is a tool that provides a reasonable trade-off between the accuracy of unsupervised heterogeneity discovery and the consistency of obtained clusters. DiviK appears to be scalable and robust enough to cope with both small- and large-scale MSI data. Moreover, it is also insensitive to modifications of the preprocessing pipeline. Compared with other configurations already proven effective for MSI, we showed that DiviK might be feasible for real-world applications when applied to solve a complex multi-dimensional problem. DiviK provides researchers with additional insight into which features were the most important for the differentiation of a specific region, which could be a subject of an even more comprehensive analysis, crucial for doctors to understand the underlying phenomena.

One could easily extend the DiviK framework to support other kinds of -omics data, with slight adjustments to the similarity measure and local filtering schema. The heterogeneity investigations based on epigenetic, transcriptomic, proteomic and metabolomic profiles require no modifications to the algorithm. Applying the DiviK algorithm to genomics data (GWAS-type) will require the application of a data-specific similarity measure and careful selection of the characteristic for the local feature filtering scheme. Finally, it is worth noting that the DiviK algorithm does not take advantage of any specific characteristics of the MSI data. Therefore, one could apply DiviK as a generic clustering algorithm, useful for grouping images after appropriate embedding (including medical images) or any other kind of tabular data.

\section*{Availability and requirements}

Project name: DiviK
\\
Project home page: \url{https://github.com/gmrukwa/divik/}
\\
Operating system(s): Linux, Windows, Mac
\\
Programming language: Python ($\geq$3.6)
\\
License: Apache 2.0
\\
Any restrictions to use by non-academics: none


\begin{backmatter}

\section*{List of abbreviations}

cdf: cumulative distribution function;
CLI: Command-Line Interface;
DiviK: Divisive intelligent K-Means;
GMM: Gaussian Mixture Modeling;
HDDC: high-dimensional data clustering;
MSI: Mass Spectrometry Imaging;
OSCC: Oral Squamous Cell Carcinoma;
PCA: Principal Components Analysis;
ROI: Region of Interest;
ToF: Time-of-Flight;
UMAP: Universal Manifold Approximation and Projection;

\section*{Declarations}
\subsection*{Ethics approval and consent to participate}

Not applicable.

\subsection*{Consent for publication}

Not applicable.

\subsection*{Availability of data and materials}

All the data is available from their original sources.

\subsection*{Competing interests}

The authors declare that they have no competing interests.

\subsection*{Funding}
This project was financially supported by NCBiR grant AIDA no. I029/17-POWR.03.02.00-IP.08-00-DOK/17 (GM) and NCN grant BITIMS no. UMO-2015/19/B/ST6/01736 (GM, JP).

\subsection*{Authors' contributions}

JP and GM designed the DiviK core algorithm. GM implemented the scripts, performed the tests, and developed the package for PyPI submission. JP and GM wrote the manuscript. JP critically revised the work. All authors read and approved the final manuscript.

\subsection*{Acknowledgements}

We would like to thank Piotr Wid\l{}ak, Monika Pietrowska, Marta Gawin and Mykola Chekan from Maria Sk\l{}odowska-Curie National Research Institute of Oncology, Gliwice branch, for providing the necessary biological and MSI data acquisition background.

Finally, we would like to thank Katarzyna Fr\k{a}tczak for help in data preprocessing.


\bibliographystyle{vancouver} 
\bibliography{bmc_article}      


\end{backmatter}
\end{document}